\documentstyle[preprint,aps]{revtex}
\tightenlines
\begin{document}
\draft
\setlength{\unitlength}{0.02mm}

\title{TWO-STRUCTURE FRAMEWORK FOR HAMILTONIAN DYNAMICAL SYSTEMS
\thanks{Work supported by the National Foundation for
Science  Researches, Contract~No.~F610/97.}}
\author{A. Petrov}
\address{Institute for Nuclear Research and Nuclear Energy,
         72 Tzarigradsko Chaussee,
         1784 Sofia,
         Bulgaria }
\date{\today}
\maketitle
\begin{abstract}
The Lie product and the order relation are viewed as defining structures
for Hamiltonian dynamical systems. Their admissible combinations are singled
out by the requirement that the group of the Lie automorphisms be contained
in the group of the order automorphisms (Lie algebras with invariant cones). 
Taking advantage of the reciprocal independence of the relevant structures,  
the inclusion relation between the two automorphism groups can be reversed;
a procedure which leads to an entirely new formal language (ordered linear 
spaces with invariant Lie products). Presumably it offers an alternative
description for quantum systems, radically different from the conventional 
algebraic models. 
\end{abstract}
{}~\\
\pacs{03.65.Bz\\ 03.65.Fd\\ 03.65.Ca}


\newpage
\section{ Objectives}
\noindent We aim at expanding the scope of the Hamiltonian formalism
as much as possible by appropriate mathematical definitions for basic concepts
like variables, states etc. The project is motivated partly by the situation
in the quantum theory where increasing pressure is felt to go beyond the limits
of the non-commutative associative algebras. One expects that the 
non-commutative associative multiplication could be replaced by weaker
mathematical structures without affecting too much the physical content
of the theory. We propose a pair of structures that promise to get the 
job done.

\section{ Lie algebras with invariant cones}
Let us name at once the structures
we are going to use: Lie algebras and order relations. We view the space 
of the variables of the Hamiltonian systems as ordered Lie algebras, or 
(which is the same thing) as Lie algebras with a fixed (positive) cone.

The Lie algebra structure is inherent in all the known classes of 
Hamiltonian systems. On the other hand, given any (abstract) Lie algebra 
with a distinguished one-parameter group of inner Lie automorphisms, we
can speak of a "Hamiltonian" and "Hamiltonian equation of motion". We have
good reasons for taking the Lie algebra structure for granted. The serious
problem is the choice of additional structures, the Lie algebra alone being
far from sufficient (and the quotation marks above more than justified).

The Lie algebra in itself is alien to the concept of states of a physical 
system. The states are usually thought of as positive linear functionals 
over the space of the variables with the implication that we know how 
the positivity is to be defined. Sure enough we know: the space of variables 
itself must be equipped with a positive cone. The point is that the Lie 
algebra structure does not generate non-trivial cones for an obvious reason --
the Lie product is antisymmetric and therefore all Lie squares are equal to
zero. The standard ordering of the associative algebras through the cone of  
the algebraic squares is not applicable to the Lie algebra context. 
Nonetheless, there are privileged candidates for cones turning (at least 
some of) the Lie algebras into ordered Lie algebras: the invariant cones 
which by definition transform onto themselves under the action of the
group of the Lie algebra automorphisms. What makes them interesting is a
sort of hereditary property -- any algebraically generated cone is
necessarily invariant. The Lie algebras give us a major example demonstrating 
that the converse is not true -- there may be invariant cones not directly
linked to the algebraic structure. There exists a complete classification
of the invariant cones in finite dimensional Lie algebras; this problem was
solved in the 1980s, but the study of the infinite dimensional case has not
yet been initiated. The invariance property is a strong requirement even if
it is restricted (as is usually done) to the group of the inner Lie 
automorphisms; it singles out a particular class of Lie algebras and prescribes
a very specific texture to the invariant cones in them.

Finally, the positive cones in the conventional models of Hamiltonian 
mechanical systems are invariant with respect to the corresponding Lie
algebras. This is equally true for the cone of the none-negative functions
on the classical phase space (with respect to the Poisson brackets), and
the cone of the positive self-adjoint operators (with respect to the 
operator commutator). Thus, the Lie algebras with invariant cones emerge
as a sound basis for an abstract model of general Hamiltonian systems. The
invariance property relating the two (otherwise independent) structures
allows immediate physical interpretation. It simply means that the group
of the Lie automorphisms is contained in the group of the order automorphisms
which, in turn, implies that the group of the common automorphisms is 
in a sense maximal. The group of the common automorphisms is actually the 
group of the symmetries of the physical system and any model of a general
dynamical system must possess a sufficiently large symmetry group. Neither
"sufficiently large" nor "maximal" is a well-defined term and the invariance
of the cone is a property open to further discussions.

One last ingredient is needed, a variable identically (in all states) 
equal to 1, and it will be represented by an order unit. We are now ready
for our first definitions.

\newpage
{\bf General model (GM).}
The space of the variables of a Hamiltonian dynamical
system is a real Lie algebra with a positive cone and order unit, both
invariant with respect to the group of the Lie automorphisms. The states are
normed positive linear functionals over the space of the variables. The 
values of the states at the elements of the Lie algebra are identified
with the mean values of the corresponding variables. The time evolution is
represented by a one-parameter group of inner Lie automorphisms.

In what follows some notations will be helpful.

{\bf Space of variables:} $(A,A^{+},e,[.,.],\omega_{t},...)$ where $A$ is a real
linear space, $A^{+}$ is a positive cone, $e$ is an order unit, $[.,.]$ is
a Lie product, and $\omega_{t}$ is the one-parameter dynamical group. The
three dots indicate everything that must be added when we want to specify
a certain class of systems or a concrete system.

{\bf Invariance requirement:} $Aut (A,[.,.]) \subseteq Aut(A,A^{+},e)$ .

{\bf Space of states:} $(V,V^{+},K)$ where $V$ is the dual space of $A$, $V^{+}$ is
the dual cone of $A^{+}$, and the set of states $K \subset V^{+}$ satisfies
$K(e) = 1$ (hence $K$ is a base for $V^{+}$).

{\bf Dynamics:} The condition that $\omega_{t}$ is in $Inn(A,[.,.])$ implies
the existence of a Hamiltonian and a Hamiltonian equation of motion.

The definitions in (GM) are extracted from the usual Hamiltonian type
dynamical models but we have discarded everything that reflects the
specific character of the classical or quantum systems. Our ultimate program
assumes that the further specification of (GM) -- up to describing concrete
physical systems -- should be carried out without leaving the language of
the invariantly ordered Lie algebras.

\section{ Hierarchy of Hamiltonian systems}
Very little can be done in the framework
of (GM) without an advanced theory of invariant cones in infinite dimensional
Lie algebras. However, some developments in the theory of ordered linear
spaces can be immediately incorporated into (GM) in order to bring the model
closer to the known algebraic descriptions.

The presence of order unit $e$ in $A^{+}$ and base $K$ in $V^{+}$ suggests the
first step towards stronger formulations.

{\bf Property (P1).} The space $(A,A^{+},e)$ is an order-unit space (and therefore
$(V,V^{+},K)$ is a base-norm space).

The terms "order-unit space" and "base-norm space" imply a little bit more
than just existence of order unit or base, respectively. The order-unit
space and the base-norm space are Banach spaces with norms determined through
the order. The transition from (GM) to (GM)+(P1) essentially means that we
choose to deal with bounded variables. A dual pair of an order-unit space
and a base-norm space has been widely discussed in the early 1970s as a 
general framework for statistical physical systems.

The boom of the theory of ordered linear spaces in the 1970s culminated in a
series of results with a key role for our purposes. It was shown that a 
certain class of order-unit spaces carry a complete spectral theory, a 
far-reaching generalization of the operator or algebraic spectral theory.
The additional properties sufficient for the existence of spectral theory
are usually introduced as "spectral duality".

{\bf Property (P2).} The spaces $(A,A^{+},e)$ and $(V,V^{+},K)$ are an order-unit
space and a base-norm space in spectral duality.

With (GM)+(P2) we reach a level of completeness which roughly corresponds
to the Von Neumann algebra models, avoiding any reference to associative
algebraic structures. The elimination of the associative algebras is not
fictitious as one might think if one remembers that every Lie algebra can
be embedded into an associative algebra. The enveloping associative algebra
is irrelevant in our context since it fails to respect the presence of the
second structure -- the invariant order relation. At the level of (GM)+(P2)
the variables are genuine random variables on Boolean event spaces, the 
events themselves related to the corresponding spectral resolutions.

The model (GM)+(P2) splits further into two categories of Hamiltonian
systems that can be given the names "classical" and "quantum", respectively.
The defining property of the classical systems is a lattice order in
$(A,A^{+},e)$ while the antilattice order can be used as a characteristic
of the quantum systems. The formulation (GM)+(P2) itself is actually the
first unified description of classical and quantum systems which is reasonably
complete and satisfies some natural aesthetic requirements.

Notice, however, that this method of specifying the classical and quantum
systems is ineffective outside (GM)+(P2); it does not work in (GM)+(P1), 
let alone in (GM).

\section{ Ordered linear spaces with invariant Lie products, or behind the
Looking-Glass}
At the very beginning of Section II there is an element of
arbitrariness that virtually predetermines (GM). There are no compelling
reasons to start with a Lie algebra and ask which is the best way of 
introducing order relations into it. There is another option that must
not be ignored -- to start with an ordered linear space and look for Lie
products consistent with the order relation. The order itself does not imply
Lie algebra structure and we have no choice but to resort again to invariance
requirements in terms of the two automorphism groups. This time, however, the
inclusion relation between them is reversed -- the group of the order
automorphisms should be contained in the group of the Lie automorphisms. 
Instead of Lie algebras with invariant cones, we introduce ordered linear
spaces with invariant Lie products. 

Treating order and Lie product on an equal footing, we can simply say that
there are two extreme cases when the group of their common automorphisms
is maximal: either $Aut(A,[.,.]) \subseteq Aut(A,A^{+},e)$ (invariance of 
the cone or, briefly, C-invariance), or $Aut(A,[.,.]) \supset 
Aut(A,A^{+},e)$ (invariance of the Lie product or, briefly, L-invariance). The
C-invariance is representative of the ordinary Hamiltonian systems as they
are described by (GM) (as it is, the C-invariance should be slightly relaxed
replacing $Aut(A,[.,.])$ by $Inn(A,[.,.])$). The L-invariance is something
new and a series of questions arise. Do L-invariant systems exist? If so, 
what can we say about their behaviour? Are they classical or quantum systems
or something else? At present no definite answers are possible. We argue
that the most plausible answers are: L-invariant Hamiltonian systems 
exist, they reveal unmistakably quantum behaviour and -- this is the crucial
point -- they are likely to offer an alternative description for what we now
regard as quantum physical systems.

The distinguishing characteristic of the hypothetical L-invariant counterpart
of (GM) is the appearance of a family of cones with a common order unit; we
have to deal with many-ordered Lie algebra $(A,\varphi A^{+},e,[.,.],
\omega_{t},...)$ where $A^{+}$ is the original cone and $\varphi$ is in the 
larger group $Aut(A,[.,.])$. Suppose now that $A^{+}$ is a lattice cone and
$(A,\varphi A^{+},e)$ is a family of vector lattices with a common order 
unit. The many-ordered Lie algebra transforms then into Lie algebra with 
many-valued commutative associative multiplication (the lattice order induces
a unique associative algebraic structure which is necessarily commutative, the
cone of the algebraic squares coinciding with the original lattice cone, and
the algebraic unit coinciding with the order unit). In more general 
(non-lattice) cases no associative algebraic structure is implied but the
many-orderedness still persists.

It is the many-orderedness of the space of the variables that suggests quantum
behaviour of the L-invariant systems. In particular, the uncertainty 
relations -- the trade-mark of the quantum systems -- are inevitable whenever
we speak of states over a many-ordered space of variables. Let us turn
again to Lie algebras ordered by a family of lattice cones with common
order unit, the L-invariant counterpart of the space of classical 
variables. Each cone of the family possesses its own dual cone with a simplex 
$\varphi'K$ as a base. The only contender for the set of states is the
intersection of the sets $\varphi'K$. Thereby the extreme points of $K$ and all
its images $\varphi'K$ (the classical pure states) lose their status of
states because they fail to satisfy the stronger positivity requirements. This
process destroys the simplicial geometry of the original set $K$ and invariably
results in uncertainty relations for some pairs of variables.

Thus, when we reformulate the hierarchy of the Hamiltonian models from
Section III to make them fit the L-invariance what we do is increasing the
number of the variables (they depend now on additional parameters describing
the family of cones) and reducing the set of states. Both changes indicate
essentially quantum nature of the L-invariant systems. The temptation is 
strong to introduce the L-invariant systems -- and in particular the vector
lattices with invariant Lie products -- as the quantum counterpart of the
classical systems (Lie algebras with invariant lattice cones according to
the classical version of (GM)+(P2)). The classical and quantum systems would
then emerge as mirror images of one another, the transition between them
being reduced to reversing the inclusion relation between the two automorphism
groups.

To this end, however, we have to remove a serious obstacle: the L-invariant
definitions are in a direct conflict with the present-day quantum theory
which is based on a single invariant non-lattice (in fact antilattice) cone
in the space of the variables (properties adopted by the quantum version of
(GM)+(P2)). This is a major challenge to the viability of our project. What we
are able to do now is singling out a class of L-invariant systems that could
be made consistent with the conventional quantum models. Our starting point
is the awareness that the L-invariant lattice model and the conventional
(C-invariant,non-lattice) quantum theory cannot be considered equally
complete if they are meant to refer to the same kind of physical objects. One
of them must be a coarsened, factorized picture of the other. Which one?

Let us look at the problem from a formal standpoint. Constructing L-invariant
models, we combine lattice order with invariant Lie product but we need not
stop there, we may go on and ask whether the resulting Lie algebra, in 
turn, admits invariant cones. If it does, beside the original L-invariant model
naturally appears a C-invariant model (possibly more than one). The transition
to the derivative (presumably non-lattice) C-invariant model is a sort of
factorization effacing some properties of the original L-invariant 
structure. In particular, the many-valued commutative multiplication in the
space of the variables is irretrievably lost.

Thus, the C-invariant language can be regarded as providing a simplified
description of the hypothetical richer L-invariant quantum models. We will
take the risk to attach the label "quantum" to the factorizable L-invariant
lattice system and regard the C-invariant quantum version of (GM)+(P2) (and
hence all operator or other conventional algebraic models) as approximations
ignoring the many-valued commutative multiplication. The conventional 
non-commutativity (when it exists at all) is nothing but a distorted 
reminiscence of the lost many-valued commutativity. Much stronger but still
plausible hypothesis is that the factorizations of the L-invariant quantum
systems are actually LC-invariant, the two automorphism groups essentially
coinciding. There is some evidence, indeed, that the typically quantum 
(factor-like) associative algebraic models exhibit such two-sided invariance.

Clearly, this classification scheme can be extended to the whole hierarchy of
Hamiltonian systems: the C-invariant systems are called classical, the
(factorizable) L-invariant systems are called quantum, and all the differences
between then can be traced to the different direction of the inclusion relation
between the two relevant automorphism groups. To avoid psychological 
difficulties, the LC-invariant systems may be separated into a special category
encompassing the factorizations of the newly introduced quantum systems (with
many-orderedness deleted). From practical point of view, the striking 
conclusion is that the standard operator quantum formalism and its algebraic
generalizations take an intermediate position -- no longer classical but not
fully quantum.

Like Lewis Carroll's little Alice, we yielded to the temptation
to reverse the invariance requirement and go through the Looking-Glass
into the dreamland of the L-invariant systems. It is uncharted virgin
territory; the L-invariant systems are legitimate mathematical objects -- this
is practically everything we know about them with certainty.

\section{ Concluding remarks}
The framework delineated by (GM) and its L-invariant
counterparts seems to exhaust the potential of the Hamiltonian formalism
for defining dynamical systems in terms of variables and states. The 
mathematical prerequisites for implementing such a program are far from
complete. The invariant cones in infinite dimensional Lie algebras and the
ordered linear spaces with invariant Lie products will remain on the agenda
for the next few decades. The program itself adds weight to the suspicion
that the non-commutative associative algebraic structure is a fallacious
beacon in the search for quantum extensions of the classical theory. At the
best it appears to be unnecessarily restrictive, and at the worst it stops
halfway between the classical and quantum systems with no chance to move
a step further without radical remodeling. Is a major part of the road to
a satisfactory quantum theory still lying ahead?

\section{ Comment on the sources}
The study of invariant cones in Lie algebras is
initiated by Vinberg \cite{1} and Paneitz \cite{2}. Hilgert and Hofmann
achieve exhaustive solutions for arbitrary finite dimensional Lie algebras
\cite {3},\cite{4}. A pair of order-unit space and base-norm space as a
suitable tool describing statistical physical systems is discussed by Davies
and Lewis \cite{5} and Edwards \cite{6}. The main contribution to the
spectral theory in the context of ordered linear spaces is made by Alfsen and 
Shultz \cite{7}. Let us remark that terms like "non-commutative spectral
theory" are misleading -- the spectral duality does not presuppose associative
algebraic structures (commutative or not). Other versions are proposed by 
Abbati and Mani\`a \cite{8} and Riedel \cite{9}. They all are
generalizations of the spectral theory for vector lattices developed by
Freudenthal \cite{10}. The antilattice geometry of the cone of the positive
self-adjoint operators is established by Kadison \cite{11}. The combination
of the geometric spectral theory of Alfsen and Shultz with the theory of 
invariant cones in Lie algebras was first propounded by the author in 
\cite{12}. The L-invariant extensions appeared in \cite{13},\cite{14}.


\end{document}